\documentclass[10pt,twocolumn,aps,pra,groupedaddress,superscriptaddress,amsmath,amssymb,showpacs,english]{revtex4-2}

\usepackage{graphicx}
\usepackage{dcolumn}
\usepackage{bm}
\usepackage[T1]{fontenc}
\usepackage{textcomp}
\usepackage{mathrsfs}
\usepackage{amsmath}
\usepackage{bbm}
\usepackage[normalem]{ulem}
\usepackage{amsbsy}
\usepackage{braket}
\usepackage{amstext}
\usepackage{amssymb}
\usepackage{physics}
\usepackage{setspace}
\usepackage{esint}
\usepackage{xcolor,colortbl}
\usepackage[colorlinks,citecolor=blue,urlcolor=blue,linkcolor=blue,hypertexnames=true]{hyperref} 
\usepackage{babel}
\usepackage{booktabs}
\usepackage{tikz}
\usepackage{mciteplus}
\begin{document}

\title{Synchronization in a dissipative quantum many-body system}

\author{Bar\i\c{s} \c{C}akmak}
\email{cakmakb@farmingdale.edu}
\affiliation{Department of Physics, Farmingdale State College—SUNY, Farmingdale, NY 11735, USA}

\author{K\"{u}bra S\"{u}mer}
\affiliation{Faculty of Engineering and Natural Sciences, Sabanci University, Tuzla, Istanbul 34956, Turkey}

\author{Steve Campbell}
\affiliation{School of Physics, University College Dublin, Belfield Dublin 4, Ireland}
\affiliation{Rinn Quantum and Centre for Quantum Engineering, Science, and Technology, University College Dublin, Belfield, Dublin 4, Ireland}

\author{G\"{o}ktu\u{g} Karpat}
\email{goktug.karpat@sabanciuniv.edu}
\affiliation{Faculty of Engineering and Natural Sciences, Sabanci University, Tuzla, Istanbul 34956, Turkey}

\date{\today}

\begin{abstract}

We study synchronization in the XX qubit chain subject to local or multi-local
amplitude-damping noise. We show that generic stable
synchronization of the edge qubits occurs \textit{if and only if} the
decoherence-free subspace (DFS) supports exactly one single-excitation
eigenstate. Analyzing the DFS structure of the model, we show
that it is completely determined by a simple number-theoretic function involving
the noise sites and the chain length. We further show that this same
condition also guarantees constant asymptotic entanglement between the edge
qubits, so that generic stable synchronization and constant asymptotic
entanglement necessarily coexist. By contrast, when the DFS supports multiple
single-excitation eigenstates, synchronization becomes initial state dependent
and may be entirely absent, even though stable oscillatory entanglement can
persist indefinitely. Our analysis can be extended to any
excitation-conserving qubit chain mappable to free fermions and is directly
testable on platforms with site-selective dissipation.

\end{abstract}

\maketitle

\textit{Introduction.}-- From the flashing of fireflies~\cite{Buck1968} and the rhythmic firing of cardiac pacemaker cells~\cite{Mirollo1990} to the coherent motion of coupled metronomes~\cite{Pantaleone2002} and the phase locking of power grids~\cite{Filatrella2008}, synchronization is a universal manifestation of collective dynamical order across nature and technology~\cite{Blekhman1988, Pikovsky2001, Balanov2009, Strogatz2001, Boccaletti2002, Acebron2005}. Over the last decade, it has also emerged as a central subject in quantum physics, where both forced synchronization under external driving and spontaneous synchronization arising solely from mutual coupling between systems have been explored. Quantum synchronization has been identified in systems as diverse as quantum van der Pol oscillators, harmonic oscillators, optomechanical and nanomechanical systems, trapped ions, superconducting circuits and spin systems~\cite{Mari2013, Li2017, Galve2017, Roulet2018a, Giorgi2019,Zhirov2008, Zhirov2009, Lee2013, Walter2014, Sonar2018, Goychuk2006, Giorgi2012, Manzano2013, Manzano2013a,Benedetti2016,Lee2014,Walter2015,Heinrich2011,Ludwig2013,Hush2015,Cabot2019,Orth2010,Giorgi2013,Bellomo2017,Giorgi2016,Karpat2020,Karpat2019,Karpat2021,Li2023a,Schmolke2022,Xiao2023,Eshaqi-sani2020, Qiao2020, Li2023, Li2022, Wachtler2023, Cattaneo2021, Cabot2021, Buca2022,Impens2023,Sterba2023,Kalit2021,Parra-Lopez2020, Tindall2020, Murtadho2023,Nadolny2023, Solanki2023, Razzoli2024, Kehrer2024, Wachtler2024, Solanki2024, Schmolke2024,Nadolny2025, Wachtler2026}, with experimental realizations demonstrated across multiple platforms~\cite{Laskar2020,Krithika2022,Zhang2023,Koppenhofer2020,Li2025,Tao2025}.

In open quantum systems, spontaneous synchronization is shaped by the interplay between coherent dynamics and environmental decay. When the decay rates of different eigenmodes are well separated, transient synchronization can appear, with one or several slowly decaying modes dominating the dynamics over a long intermediate time window before eventually fading away~\cite{Giorgi2019}. By contrast, stable synchronization requires the existence of a decoherence-free subspace (DFS)~\cite{Lidar2003,Buca2022}, i.e., a noiseless sector of the dynamics that remains decoupled from the environment and thus can sustain persistent oscillations indefinitely. This also leads to non-stationary phenomena in the steady states of open quantum systems, such as dissipative time-crystalline behavior~\cite{Buca2019,Guarnieri2022,Campbell2025}.

Quantum synchronization in an open quantum many-body setting was recently predicted by Schmolke and Lutz~\cite{Schmolke2022} for a chain of qubits subjected to local Gaussian white noise, using a perturbative approach in Liouville space, and  was subsequently observed experimentally in superconducting transmon circuits~\cite{Tao2025}. However, emergence of stable synchronization in dissipative qubit chains under amplitude-damping (AD) noise, and more importantly its connection to the DFS structure, remains unexplored. In this manuscript, we study synchronization by directly analyzing the DFS structure of the XX qubit chain subject to local or multi-local AD noise. We first demonstrate that the DFS structure is entirely fixed by a simple number-theoretic function, namely, the greatest common divisor of the noise sites and the chain length. Then, we derive a closed-form expression for the expectation values of local qubit observables restricted to the DFS and establish a necessary and sufficient condition for generic stable synchronization of the edge qubits. More precisely, we prove that generic stable synchronization occurs \textit{if and only if} the DFS supports exactly one single-excitation eigenstate. We also show that generic stable synchronization and constant asymptotic entanglement between the edge qubits always coexist, as both are enforced by exactly the same condition on the number of DFS eigenstates. However, when synchronization loses its generic character and becomes initial state dependent, oscillatory entanglement can still persist indefinitely without synchronization ever being established.

\textit{The model and its DFS structure}-- We focus on the XX chain of qubits subjected to local or multi-local AD noise. Under certain physically motivated assumptions, the dynamics of this open many-body system is described by the GKSL equation~\cite{Lindblad1976,Gorini1976,BreuerPetruccione}
\begin{equation}\label{eq:me}
    \dot{\rho}=-i\left[H, \rho\right]+\mathcal{D}(\rho),
\end{equation}
where $\mathcal{D}(\rho)=\sum_i \gamma_i[L_i\rho L^{\dagger}_i-\tfrac{1}{2}\{L^{\dagger}_iL_i,\rho\}]$ is the dissipator with $\gamma_i$ being the strength of the coupling to the environment, $L_i$ are the jump operators, and $H=H_S+H_{\text{int}}$ includes the self-Hamiltonians of the qubits making up the chain and the pairwise coupling between them, respectively. Specifically, the Hamiltonian that governs the unitary part of the evolution is given by 
\begin{align}\label{HXX}
H&=-\frac{\omega}{2}\sum^{N}_{i=1}\sigma_z^{(i)}+\frac{J}{2}\sum^{N-1}_{i=1}\left(\sigma_{x}^{(i)}\sigma_{x}^{(i+1)}+\sigma_{y}^{(i)}\sigma_{y}^{(i+1)}\right)
\end{align}
where $\sigma_{x},\sigma_{y}$ and $\sigma_{z}$ are the usual Pauli operators, and $N$ denotes the number of qubits in the chain. Note that $H$ conserves the number of excitations and does not couple excitation sectors with different numbers of excitations. Consequently, we start our DFS analysis by focusing on the single-excitation subspace of $H$. 

We will denote the basis vectors of the single-excitation subspace with $\ket{j}=\ket{0\cdots1_{j}\cdots0}$ for $j=1,\dots, N$. Using  $\sigma_{\pm}=(\sigma_x\mp i\sigma_y)/2$ and setting the ground state energy to
$E_0=0$, it is possible to express the single-excitation sector of the full Hamiltonian $H$ given in Eq.~(\ref{HXX}) in this basis as a nearest-neighbor hopping Hamiltonian, that is, $H^{(1)}=\omega\sum_{j=1}^N\ket{j}\bra{j}+J\sum_{j=1}^{N-1}\big(\ket{j}\bra{j+1}+\ket{j+1}\bra{j}\big),$
which is a Toeplitz tridiagonal matrix. The eigenvalues of such matrices are given by $E_n=\omega+2J\cos(\frac{n\pi}{N+1})$ and the corresponding eigenvectors read~\cite{Takahashi1999}
\begin{equation}\label{eq:amplitudes}
|\phi_n\rangle=\sum_{j=1}^N \phi_n(j)\,|j\rangle,\quad
\phi_n(j)=\sqrt{\tfrac{2}{N+1}}\sin\bigl(\tfrac{nj\pi}{N+1}\bigr),
\end{equation}
where $n=1,\dots,N$ and the amplitudes satisfy the parity symmetry relation $\phi_n(N + 1 - j) = (-1)^{n+1}\phi_n(j)$.
Let us now focus on the DFS structure of the XX chain under local AD noise in the single-excitation subspace. A subspace is called decoherence-free if (i) every state $\ket{\psi}$ in it is a simultaneous eigenstate of all jump operators, $L_i\ket{\psi}=c_i\ket{\psi}$, with $c_i$ being independent of the state $\ket{\psi}$ within this subspace, and (ii) the subspace is invariant under $H$ so that the unitary dynamics does not couple it to the decohering sector~\cite{Zanardi1997,Lidar1998,Knill2000,Lidar2003}. For AD noise on the site $m$, $L=\sigma^{(m)}_-$ acts on the single-excitation basis states as $\sigma^{(m)}_-\ket{j}=0$, for $j\neq m$. Furthermore, $\sigma^{(m)}_-\ket{m}=\ket{G}$, where $\ket{G}\equiv \ket{0}^{\otimes N}$ denotes the ground state. Physically, this means that all single-excitation basis states with no excitation on the noise site are unaffected by the noise and thus constitute the set of dark basis states. By contrast, $\ket{m}$ is the only bright basis state, since it is coupled by dissipation to the ground state and thus leaks excitation out of the single-excitation sector.

In principle, the linear combinations of the dark basis states are decoupled from dissipation due to the AD jump operator. However, to belong to the DFS, they must also remain in the dark sector under unitary dynamics. Since the XX Hamiltonian generates nearest-neighbor hopping, a generic superposition of dark basis states evolves into a state with non-zero amplitude on the noisy site and thus leaks out of the dark sector. The single-excitation states that remain dark for all times are thus the eigenstates of $H$ with zero amplitude on the noise site $m$. This requires that $\phi_n(m)=0$, leading to $\frac{nm}{N+1}\in\mathbb{Z}$, which in turn implies that the number of single-excitation DFS states is $r=g-1$, where $g=\text{gcd}(m, N+1)$, and the eigenstate labels that belong to the DFS are given by $n_l=l(N+1)/g$ with $l=1, \dots, g-1$~\cite{sm}. Here, gcd denotes the greatest common divisor function. This will allow us to fully characterize the DFS for an XX chain of arbitrary length subject to local AD noise at arbitrary sites, and thus constitutes the first main result of this manuscript.

More generally, let $\{\ket{D_{\alpha}}\}$, where $\alpha\in\{n_l\}$, denote the set of single-excitation DFS states with coefficients $\phi_{\alpha}(j)=\langle j|D_{\alpha}\rangle$. Note that the DFS also includes the zero-excitation state $\ket{G}$. Since the XX chain Hamiltonian can be mapped to a free-fermion model~\cite{Lieb1961,Takahashi1999}, the multi-excitation DFS sectors can be obtained from distinct single-excitation DFS eigenmodes. Particularly, for  $\alpha\neq\beta$, two-excitation DFS states are constructed as
\begin{align}\label{slater}
    &\ket{D_{\alpha\beta}}=\sum_{1\leq i<j\leq N}\Psi_{\alpha\beta}(i, j)\ket{i, j},
\end{align}
where $\Psi_{\alpha\beta}(i, j)=\phi_{\alpha}(i)\phi_{\beta}(j)-\phi_{\alpha}(j)\phi_{\beta}(i)$, with additive energies, that is, $E_{D_{\alpha\beta}}=E_{D_{\alpha}}+E_{D_\beta}$. All higher-excitation DFS states are obtained as $k\times k$ Slater determinants built from $k$ distinct single-excitation DFS eigenmodes. Thus, if the single-excitation DFS contains $r$ eigenmodes, the $k$-excitation DFS sector has dimension $\binom{r}{k}$, where $k=0,1,\dots,r$. Then, $r=\text{gcd}(m,N+1)-1$ not only determines the number of single-excitation DFS states, but also identifies all the higher-excitation DFS sectors. Hence, the full DFS structure of the XX chain under AD noise at site $m$ is completely characterized.

The above result extends naturally to the case of multi-local AD noise. When the noise acts on sites $m_1,\dots,m_q$, a single-excitation eigenmode belongs to the DFS if it has zero amplitude on each noise site, i.e., $\phi_{\alpha}(m_a)=0$ for all $a=1,\dots,q$. This leads to
\begin{equation}
r=g-1, \qquad g=\gcd(m_1,\dots,m_q,N+1),
\end{equation}
so that, as in the local AD noise case, both the number of single-excitation DFS states and the full DFS structure are completely determined by a gcd condition~\cite{sm}.

Our DFS construction identifies invariant state vectors in Hilbert space. A more general Liouville space approach can admit additional non-decaying operators~\cite{Lidar2003}. For the present model, however, the two approaches coincide, since nearest-neighbor hopping moves excitations across the chain, and any component that reaches a noise site is removed by AD noise. Thus, all non-decaying density operators have support entirely within the subspace spanned by the DFS eigenstates identified above.

Lastly, the outlined DFS structure cannot be preserved for finite temperature AD noise, which will introduce an additional jump operator $\sigma^{(m)}_+$ alongside $\sigma^{(m)}_-$. All DFS eigenstates identified above have, by construction, vanishing amplitude at the noise site $m$. The jump operator $\sigma^{(m)}_+$ creates an excitation precisely at that site, moving these states out of the DFS. Consequently, no non-trivial DFS exists under generalized local AD noise~\cite{Campbell2025}.

We emphasize that the construction above is not specific to the uniform XX chain. In fact, it relies only on two structural properties of the model: (i) the Hamiltonian conserves the total number of excitations, and (ii) it can be mapped to non-interacting fermions. For any qubit chain having these properties, the single-excitation DFS states are the eigenmodes with vanishing amplitude on every noise site, and the multi-excitation DFS states are their Slater determinants with additive energies, and the synchronization analysis developed below carries over without modification. The gcd criterion is the manifestation of this dark-mode counting for the uniform chain, where the eigenstate amplitudes are of sine form.

\textit{Conditions for generic synchronization.}-- In this part, we establish the conditions under which stable synchronization is achieved between the two edge qubits, and determine when this synchronization is generic. By stable synchronization, we mean sustained non-decaying oscillations in the expectation values, as expected from a non-stationary steady state. The expectation value of interest in our treatment is $\langle \sigma_{x}^{(k)}(t) \rangle = \text{Tr} [ \rho(t) \sigma_{x}^{(k)} ]$. Subsequent to the transient dynamics, the asymptotic state of the system is confined to the DFS, where the dynamics is purely unitary, which yields the closed form expression~\cite{sm}
\begin{equation} \label{eq:genexpval}
\langle \sigma_x^{(k)}(t)\rangle_{\text{DFS}}
= 2\mathrm{Re}
\sum_{\substack{\mu,\nu\in\text{DFS}}}
\rho^{\infty}_{\nu\mu}
X_{\mu \nu}^{(k)}
e^{-i(E_\nu-E_\mu)t}.
\end{equation}
In the above equation, the summation runs over pairs of DFS eigenstates $\ket{\mu}$ and $\ket{\nu}$, whose excitation numbers differ by one, i.e. $\mathcal{N}_{\nu}=\mathcal{N}_{\mu}+1$, since $\sigma_x=\sigma_++\sigma_-$, and the matrix element $\rho^{\infty}_{\nu\mu}$ is the coherence between $\ket{\mu}$ and $\ket{\nu}$ in the density operator $\rho^{\infty}$ at the onset of the unitary regime, after all bright components have decayed. $X_{\mu \nu}^{(k)}=\bra{\mu}\sigma_x^{(k)}\ket{\nu}$ encodes whether flipping the qubit at site $k$ can connect the two DFS states $\ket{\mu}$ and $\ket{\nu}$. Hence, each allowed DFS transition contributes an oscillatory component at frequency $\omega_{\nu\mu}=E_\nu-E_\mu$. If multiple distinct transitions survive, the dynamics exhibits multi-frequency oscillations, whereas a single surviving transition leads to oscillations at a single frequency. 

We define generic stable synchronization between the edge qubits by the condition
\begin{equation}\label{eq:synccriterion}
    \langle \sigma_x^{(1)}(t) \rangle_{\text{DFS}}
    =
    C\,\langle \sigma_x^{(N)}(t) \rangle_{\text{DFS}},
\end{equation}
for all $t$, once the dynamics is confined to the DFS, where $C$ is a real constant, and neither asymptotic edge expectation value vanishes at all times. Even though this criterion excludes arbitrary locked phases, it is
well suited to the present setting, where the surviving asymptotic solutions
are exactly in phase or in antiphase, consistent with the parity symmetry of
the eigenstate amplitudes below Eq.~(\ref{eq:amplitudes}). Since generic synchronization requires Eq.~(\ref{eq:synccriterion}) to hold for arbitrary initial states~\cite{note}, Eq.~(\ref{eq:genexpval}) implies $X_{\mu\nu}^{(1)}=CX_{\mu\nu}^{(N)}$ for all contributing DFS pairs $\mu,\nu$. \emph{If} there exists only one single-excitation DFS state, i.e. $g=2$, then $X_{GD_{\alpha}}^{(1)}=\phi_{\alpha}(1)$ and $X_{GD_{\alpha}}^{(N)}=\phi_{\alpha}(N)$, which yields $\phi_{\alpha}(1)=(-1)^{\alpha+1}\phi_{\alpha}(N)$ where the latter equality follows from the inversion symmetry of the eigenstate amplitudes given in Eq.~(\ref{eq:amplitudes}). The edge qubits are thus synchronized for odd $\alpha$ and anti-synchronized for even $\alpha$, where $\alpha=n_1=(N+1)/g=(N+1)/2$. This clearly implies that $N$ is odd and synchronization is generic. The frequency of the synchronized oscillations is dictated by the energy difference $E_{D_{\alpha}}-E_G=E_{D_{\alpha}}$. 

It can also be shown that generic synchronization occurs \emph{only if} the DFS supports one single-excitation eigenstate. In order to see this, consider the case of two single-excitation DFS states, which allows for the construction of a single two-excitation DFS state. Then, we need to account for matrix elements connecting single- and two-excitation DFS states. We focus on the matrix elements contributing to the same oscillatory term as the $X_{GD_{\alpha}}$ transition. Since $E_{D_{\alpha\beta}}-E_{D_{\beta}}\!=\!(E_{D_{\alpha}}+E_{D_{\beta}})-E_{D_{\beta}}\!=\!E_{D_{\alpha}}$ matches the frequency of the $X_{GD_{\alpha}}$ transition, those matrix elements are $X_{D_{\beta}D_{\alpha\beta}}^{(1)}$ and $X_{D_{\beta}D_{\alpha\beta}}^{(N)}$, and must thus satisfy the same sign condition. Nonetheless, it is possible to show that $X_{D_{\beta}D_{\alpha\beta}}^{(1)}=\phi_{\alpha}(1)$ and $X_{D_{\beta}D_{\alpha\beta}}^{(N)}=-\phi_{\alpha}(N)$, and therefore $\phi_{\alpha}(1)=-(-1)^{\alpha+1}\phi_{\alpha}(N)$. In fact, this is incompatible with the previously obtained relation above, $\phi_{\alpha}(1)=(-1)^{\alpha+1}\phi_{\alpha}(N)$, making it impossible for all the terms to satisfy the synchronization condition simultaneously. Consequently, generic stable synchronization emerges \emph{if and only if} there exists only one single-excitation DFS state~\cite{sm}, i.e.,
\begin{equation}\label{eq:iffcond}
g=\gcd(m_1,\dots,m_q,N+1)=2,
\end{equation}
which is the second main result of this manuscript.

The same condition also fixes the synchronization behavior at every site of the chain, that is, the amplitudes $\phi_{\alpha}(j)=\sqrt{2/(N+1)}\sin(j\pi/2)$ vanish on even sites and alternate in sign on consecutive odd sites. Thus, the odd-site qubits split into two classes that are synchronized within each class and anti-synchronized between them, consistent with the parity symmetry of the eigenstates.

\begin{figure*}[t]
    \centering
    \includegraphics[width=1.0\textwidth]{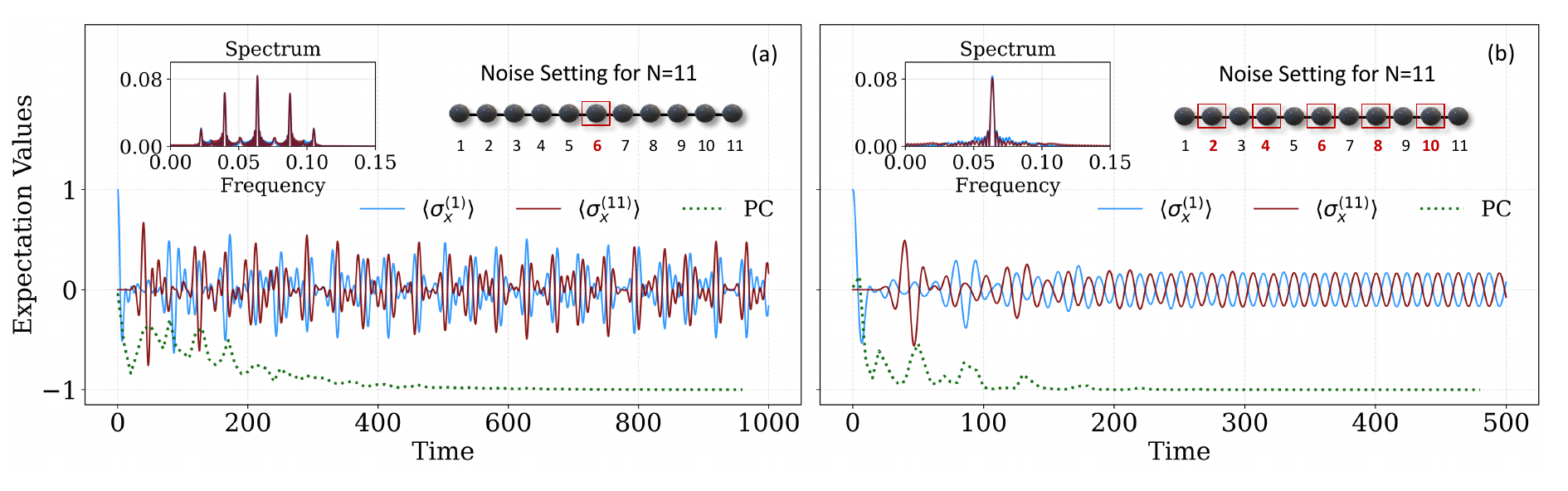}
    \caption{Stable synchronization of expectation values $\langle \sigma_x \rangle$ for the first and the last qubits of the XX chain of $N=11$ qubits. (a) Only the middle qubit at the site $m=6$ is affected by the AD noise. (b) The qubits at sites $m=2,4,6,8,10$ are affected by the AD noise having identical noise strengths $\gamma$. While the insets display the frequency spectrum of the $\langle \sigma_x \rangle$ oscillations, the dynamical establishment of stable anti-synchronization is demonstrated by the behavior of the Pearson coefficient (PC). In both plots, the initial state is $\ket{\psi}=\ket{+}_{1} \otimes \bigotimes^{11}_{k=2} \ket{0}_{k}$ and we fix the model parameters as $\gamma=0.05, J=0.15, \omega=0.4$. It is evident that while there are five distinct frequencies in (a) contributing to the anti-synchronized dynamics of the edge qubits, in (b) we observe single-frequency synchronization of the edge qubits. The anti-synchronization observed in panel (a) is state dependent rather than generic, since the chosen initial state has support only in the zero- and single-excitation sectors.}
    \label{fig:Fig1}
\end{figure*}
\textit{Case study.}-- Here, we present a concrete demonstration of our general results in a chain consisting of $N=11$ qubits, considering both local and multi-local noise cases. To explicitly track the synchronization behavior at all times during the dynamics, we adopt the well-established figure of merit for this purpose, namely the Pearson coefficient (PC), defined as $\text{PC}\!=\!\text{Cov}(x,y)/\sqrt{\text{Var}(x)\text{Var}(y)}$, where $x\!=\!\langle \sigma_x^{(1)}(t) \rangle$ and $y\!=\!\langle \sigma_x^{(N)}(t) \rangle$ in our case~\cite{Galve2017}. 

For local noise, we take the noise site to be $m\!=\!6$, which admits $r\!=\!\text{gcd}(6,12)-1=5$ single-excitation DFS states, i.e., $\{\ket{D_{2}}, \ket{D_{4}}, \ket{D_{6}}, \ket{D_{8}}, \ket{D_{10}}\}$, with corresponding energies $E_{\alpha}=\omega+\sqrt{3}J, \omega+J, \omega, \omega-J, \omega-\sqrt{3}J$, respectively. By our second main result, this rules out generic synchronization. All the same, for certain classes of initial states, the system can still sustain multi-frequency synchronization~\cite{sm}. In particular, it is clear from Eq.~(\ref{eq:genexpval}) that the frequencies of the oscillations are set by the energy difference between the DFS states that differ by a single excitation, namely $\omega_{\nu\mu}$. Recalling that the energies of the multi-excitation DFS states are determined additively from the single-excitation DFS energies, we see that the relevant frequencies are dictated only by the energy differences between the ground state and single-excitation DFS eigenstates, $\omega_{D_{\alpha}G}=E_{D_{\alpha}}-E_G=E_{D_{\alpha}}$. In Fig.~\hyperref[fig:Fig1]{\ref*{fig:Fig1}(a)}, we display the full dynamics of our model with the aforementioned noise setting for the initial state $\ket{\psi}=\ket{+}_{1} \otimes \bigotimes^{11}_{k=2} \ket{0}_{k}$ and model parameters $\gamma=0.05, J=0.15, \omega=0.4$. Following a transient dynamics, we can clearly observe the establishment of anti-synchronization between the edge qubits indicated by the settlement of PC to $-1$. This behavior can be understood by recalling the parity symmetry of the eigenstates forming the DFS, together with the fact that all $\alpha$ values are even in the present case, i.e., $\phi_{\alpha}(1)/\phi_{\alpha}(N)\!=\!-1$. Due to the same parity symmetry argument, every pair that is reflection symmetric with respect to the center of the chain is also automatically synchronized. Moreover, it is possible to see from the behavior of the $\langle \sigma_x \rangle$ values that we do not have a simple single-frequency anti-synchronization. By taking the Fourier transform of the $\langle \sigma_x \rangle$ data, we explicitly show in the upper-left inset of Fig.~\hyperref[fig:Fig1]{\ref*{fig:Fig1}(a)} that there exist in fact $5$ distinct frequencies contributing to the anti-synchronized dynamics of the edge qubits. These are exactly the predicted frequencies dictated by the single-excitation DFS energies, which, for the considered model parameters, are $\omega_{D_{\alpha}G}/2\pi=0.022, 0.040, 0.064, 0.088, 0.105$. This presents an explicit demonstration of a fully coherent multi-frequency synchronization in a controlled setting.

For multi-local noise, we extend the noise sites to include $m_a=2, 4, 6, 8, 10$, as depicted in the upper-right inset of Fig.~\hyperref[fig:Fig1]{\ref*{fig:Fig1}(b)}. Since $\text{gcd}(2, 4, 6, 8, 10, 12)=2$, this configuration reduces the size of the single-excitation DFS set to a single element, which is $\ket{D_6}$. Note that the current setting satisfies Eq.~(\ref{eq:iffcond}), therefore we observe generic synchronization with its frequency, $f=\omega/2\pi$, determined by the energy difference $E_{D_6}-E_G=\omega$. Using the same initial state and model parameters as the local noise case, from Fig.~\hyperref[fig:Fig1]{\ref*{fig:Fig1}(b)} we clearly observe the establishment of anti-synchronization of the edge qubits sustained by a single frequency that is extracted from the Fourier transform of the expectation values and displayed in the upper-left inset of the same figure. A direct comparison between Fig.~\hyperref[fig:Fig1]{\ref*{fig:Fig1}(a)} and \hyperref[fig:Fig1]{\ref*{fig:Fig1}(b)}, shows that as the number of noise sites is increased, we observe earlier establishment of synchronization. This can be understood from the fact that increased noise allows a faster decay into the DFS.

\begin{figure}[t]
    \centering
    \includegraphics[width=1.0\columnwidth]{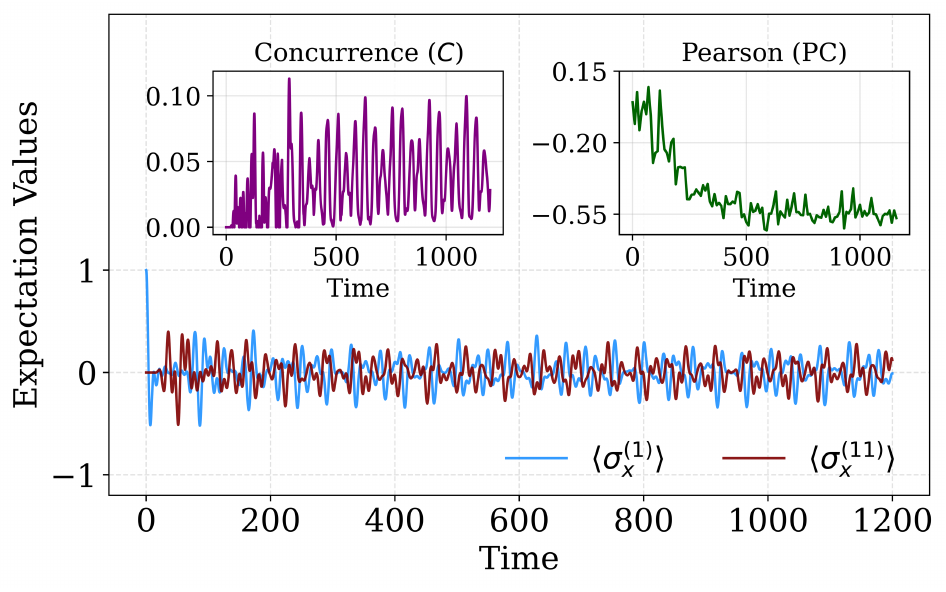}
    \caption{Asynchronous dynamics of expectation values $\langle \sigma_x \rangle$ for the first and the last qubits of the XX chain of $N=11$ qubits, where the relaxation noise is at the site $m=6$. The insets show the dynamics of entanglement and Pearson coefficient (PC) for the edge qubits. Whereas the initial state is fixed to $\ket{\psi}=\ket{+}_{1} \otimes \ket{-}_{2} \otimes  \bigotimes^{11}_{k=3} \ket{0}_{k}$, the model parameters are taken as $\gamma=0.05, J=0.15, \omega=0.4$.}
    \label{fig:Fig2}
\end{figure}

Whenever the \emph{iff} condition for generic synchronization in Eq.~(\ref{eq:iffcond}) is satisfied, which requires odd $N$, it is necessarily accompanied by a constant finite asymptotic entanglement between the edge qubits~\cite{sm}. For example, the asymptotic concurrence for the initial state considered in  Fig.~\hyperref[fig:Fig1]{\ref*{fig:Fig1}} scales as $\mathcal{C}^\infty_{(1,N)}=4/(N+1)^2$~\cite{Wootters1998}. Nevertheless, entanglement between the edge qubits can also survive in the absence of generic synchronization, and it is not necessarily constant. In Fig.~\hyperref[fig:Fig2]{\ref*{fig:Fig2}}, we consider the exact same scenario as in Fig.~\hyperref[fig:Fig1]{\ref*{fig:Fig1}(a)} but with a different initial state, $\ket{\psi}=\ket{+}_{1} \otimes \ket{-}_{2} \otimes  \bigotimes^{11}_{k=3} \ket{0}_{k}$. The PC displayed in the upper-right inset confirms synchronization is not established, yet the upper-left inset reveals oscillatory but finite concurrence between the edge qubits. While the interplay between synchronization and quantum
correlations has been examined in a number of different works~\cite{Giorgi2012,
Manzano2013,Giorgi2013,Bellomo2017,Karpat2019}, our results sharpen their relation in the present many-body setting, that is, constant asymptotic edge entanglement
necessarily accompanies generic synchronization, whereas persistent oscillatory entanglement can survive even when synchronization is completely absent. This demonstrates that, despite their coexistence when $g=2$, synchronization and entanglement are ultimately driven by different physical mechanisms in general.

\textit{Conclusion.}-- Our analysis shows that a single integer, namely the greatest common divisor of the noise sites and the chain length, fully determines the DFS structure governing the long-time coherent behavior of the dissipative XX chain under local or multi-local noise. Generic stable synchronization of the edge qubits emerges if and only if the DFS supports exactly one single-excitation eigenstate. Under the same condition, the edge qubits also develop constant asymptotic entanglement, so that both phenomena are in fact controlled by a single arithmetic condition. When the DFS supports multiple single-excitation eigenstates, synchronization loses its generic character, becomes sensitive to the initial state, and may exhibit a multi-frequency structure dictated by the single-excitation DFS spectrum. Entanglement, however, can persist indefinitely even without any synchronization being established. Our results are analytical and testable in current experimental platforms with site-selective dissipation, such as superconducting qubit arrays~\cite{Tao2025} and trapped ions~\cite{Li2025,Wachtler2025}. Since the underlying mechanism requires only that the
Hamiltonian conserves the total number of excitations and can be mapped to
non-interacting fermions, the present framework extends directly to more
general qubit chains with non-uniform couplings and non-resonant on-site energies. There, the sine wave eigenmodes no longer apply and the gcd criterion can be lost, yet the dark-mode counting that governs generic synchronization remains
exactly characterizable.

\textit{Acknowledgments.}--  B.Ç., K.S. and G.K. are  supported by the Scientific and Technological Research Council of Türkiye (TUBITAK) under Grant No. 125F434. B.Ç. is partially supported by the Farmingdale State College Provost's Office Summer Research Scholarship.  G.K. is supported by the Scientific and Technological Research Council of Türkiye (TUBITAK) through the 100th Anniversary Incentive Award. S.C. acknowledges support from the John Templeton Foundation Grant ID 63626 and from Taighde \'Eireann - Research Ireland under grant number 24/EPSRC/4121. This publication has emanated from research conducted with the financial support of Taighde \'Eireann – Research Ireland, under Grant No. 23/RC/12197 at Rinn Quantum.

\bibliography{bibliography}

\clearpage
\onecolumngrid

\newpage
\onecolumngrid
\begin{center}
    {\large \textbf{Supplementary Material: Synchronization in a dissipative quantum many-body system}} \\[0.5cm]
\end{center}

\setcounter{equation}{0}
\setcounter{section}{0}
\setcounter{figure}{0}
\setcounter{table}{0}
\setcounter{page}{1}

\renewcommand{\theequation}{S\arabic{equation}}
\renewcommand{\thefigure}{S\arabic{figure}}
\renewcommand{\thesection}{\Roman{section}}
\renewcommand{\thepage}{S\arabic{page}}

\section{Determination of the DFS Structure}

We determine the single-excitation DFS states by requiring that the eigenstate amplitudes vanish at the noise site:
\begin{equation}
\phi_n(m)=\sqrt{\frac{2}{N+1}}\sin\!\left(\frac{nm\pi}{N+1}\right)=0, \qquad \text{i.e.,} \qquad \frac{nm}{N+1}\in\mathbb{Z}.
\end{equation}
Let us first assume that $m$ and $N+1$ are coprime, that is, $g=\gcd(m,N+1)=1$. Then the condition $nm/(N+1)\in\mathbb{Z}$ implies that $N+1$ divides $n$. Since $1\leq n\leq N$, this is impossible and thus no single-excitation eigenstate has a node at site $m$, meaning that there exists no single-excitation DFS state. Next, suppose that $g=\gcd(m, N+1)\geq 2$. Writing $m=gm'$ and $N+1=gL$, the divisibility condition becomes $nm'/L\in\mathbb{Z}$. Since $\gcd(m',L)=1$ by construction, and the constraint $1\leq n\leq N$ restricts $l$ to $l=1,\ldots,g-1$ ($l=g$ would give $n=N+1$, which lies outside the allowed values of $n$), this requires $n$ to be an integer multiple of $L=(N+1)/g$, i.e.,
\begin{equation}\label{eq:DFSeigenstates}
n_l=l\,\frac{N+1}{g}, \qquad l=1,2,\ldots, g-1.
\end{equation}
 The number of single-excitation DFS states is thus given by $r=g-1=\gcd(m,N+1)-1$.

This result can be generalized for local AD noise acting on multiple sites, such as $m_1,\dots,m_q$. A single-excitation eigenmode $|\phi_n\rangle$ belongs to the DFS if and only if $\phi_n(m_a)=0$ for every $a=1,\ldots,q$, which requires
\begin{equation}
\frac{nm_a}{N+1}\in\mathbb{Z}, \qquad a=1,\ldots,q.
\end{equation}
For each site $m_a$, this constrains $n$ to be a multiple of $(N+1)/g_a$, where $g_a=\gcd(m_a,N+1)$. Since a single label $n$ must satisfy all $q$ conditions simultaneously, the allowed values are the common multiples of the integers $(N+1)/g_1,\ldots,(N+1)/g_q$. The smallest such multiple is given by the least common multiple (lcm) function, so that
\begin{equation}
n_l=l\,\mathrm{lcm}\!\left(\frac{N+1}{g_1},\ldots,\frac{N+1}{g_q}\right).
\end{equation}
As each $(N+1)/g_a$ divides $N+1$, the integers $(N+1)/g_a$ are divisors of the same number $N+1$. Consequently, we have $\mathrm{lcm}[(N+1)/g_1,\ldots,(N+1)/g_q] = (N+1)/\gcd(g_1,\ldots,g_q)$. In addition, since every common divisor of $g_1,\dots,g_q$ divides each $m_a$ and $N+1$, and conversely every common divisor of $m_1,\dots,m_q,N+1$ divides each $g_a=\gcd(m_a,N+1)$, it follows that $\gcd(g_1,\ldots,g_q)=\gcd(m_1,\ldots,m_q,N+1)\equiv g$. Hence, the allowed labels are
\begin{equation}
n_l=l\,\frac{N+1}{g}, \qquad l=1,\ldots,g-1.
\end{equation}
Then, the number of single-excitation DFS states is $r=g-1=\gcd(m_1,\ldots,m_q,N+1)-1$.

\section{Derivation of the closed-form of $\langle \sigma_x^{(k)}(t)\rangle_{\text{DFS}}$}

Here, we derive a closed-form expression for the DFS contribution to the expectation value $\langle \sigma_{x}^{(k)}(t) \rangle = \text{Tr} [ \rho(t) \sigma_{x}^{(k)} ]$. Let $\Pi_{\text{DFS}}$ be the orthogonal projector onto the DFS. Whereas the considered observable restricted to the DFS is $\sigma_{x,\text{DFS}}^{(k)} = \Pi_{\text{DFS}} \sigma_{x}^{(k)} \Pi_{\text{DFS}}$, the DFS block of the full density operator becomes $\rho_{\text{DFS}}(t) = \Pi_{\text{DFS}} \rho(t) \Pi_{\text{DFS}}$. Consequently, the DFS contribution to the expectation value becomes
\begin{align}
\langle \sigma_x^{(k)}(t)\rangle_{\text{DFS}} &= \text{Tr}\left[\rho_\text{DFS}(t) \sigma_{x,\text{DFS}}^{(k)}  \right] = \text{Tr}\left[ \rho_\text{DFS}(t)   \Pi_{\text{DFS}} \sigma_{x}^{(k)} \Pi_{\text{DFS}}  \right] \nonumber \\
 &= \text{Tr}\left[ \Pi_{\text{DFS}} \rho_\text{DFS}(t)   \Pi_{\text{DFS}} \sigma_{x}^{(k)}  \right] =\text{Tr}\left[ \rho_\text{DFS}(t) \sigma_{x}^{(k)}  \right],
\end{align}
where we first used the cyclicity of the trace and then the fact that $\rho_\text{DFS}(t) = \Pi_{\text{DFS}} \rho_\text{DFS}(t) \Pi_{\text{DFS}}$.

Let us consider an orthonormal DFS basis $\{\ket{\nu}\}_{\nu\in\text{DFS}}$ consisting of Hamiltonian eigenstates,
\begin{equation}
H\ket{\nu}=E_\nu\ket{\nu},
\end{equation}
where $H$ is the full Hamiltonian involving all excitation sectors. Then, we can write the trace over the DFS as
\begin{equation}
\langle \sigma_x^{(k)}(t)\rangle_{\text{DFS}}
=\text{Tr}\left[ \rho_\text{DFS}(t) \sigma_{x}^{(k)}  \right]=\sum_{\nu\in\text{DFS}}\bra{\nu}\rho_{\text{DFS}}(t)\sigma_x^{(k)}\ket{\nu}.
\end{equation}
In the above equation, since $\rho_{\text{DFS}}(t)$ has support only within the DFS, the trace over the full Hilbert space reduces to a sum over DFS eigenstates. Inserting  $\mathbb{I}_{\text{DFS}} = \sum_{\mu \in \text{DFS}} | \mu \rangle \langle \mu |$ between the operators $\rho_{\text{DFS}}(t)$ and $\sigma_x^{(k)}$ above, 
\begin{equation}
\langle \nu | \rho_{\text{DFS}}(t) \sigma_x^{(k)} | \nu \rangle = \sum_{\mu \in \text{DFS}} \langle \nu | \rho_{\text{DFS}}(t) | \mu \rangle \langle \mu | \sigma_x^{(k)} | \nu \rangle.
\end{equation}
Consequently, the DFS contribution to the $\sigma_x$ expectation value at site $k$ takes the form
\begin{equation} \label{expvalue1}
\langle \sigma_x^{(k)}(t)\rangle_{\text{DFS}}
=\sum_{\mu,\nu\in\text{DFS}} (\rho_{\text{DFS}}(t))_{\nu\mu}\bra{\mu}\sigma_x^{(k)}\ket{\nu},
\end{equation}
where $(\rho_{\text{DFS}}(t))_{\nu\mu}=\bra{\nu}\rho_{\text{DFS}}(t)\ket{\mu}$. In the asymptotic regime, after all bright components have decayed and dissipative feeding into the DFS has ceased, the evolution within the DFS becomes purely unitary and thus
\begin{equation} \label{rhodfst}
\rho_{\text{DFS}}(t)=e^{-iHt}\rho^{\infty}e^{+iHt},
\end{equation}
where $\rho^{\infty}$ denotes the asymptotic density operator at the onset of the purely unitary regime, which is within the DFS. It depends both on the initial state of the system $\rho(0)$ and the transient dissipative feeding into the DFS. Substituting Eq.~(\ref{rhodfst}) in $(\rho_{\text{DFS}}(t))_{\nu\mu}$ above, we end up with
\begin{equation}
(\rho_{\text{DFS}}(t))_{\nu\mu}
=\bra{\nu}e^{-iHt}\rho^{\infty}e^{+iHt}\ket{\mu} =e^{-i(E_\nu-E_\mu)t}\rho^{\infty}_{\nu\mu}.
\end{equation}
Inserting this equation back again in Eq.~(\ref{expvalue1}), we obtain
\begin{equation}
\langle \sigma_x^{(k)}(t)\rangle_{\text{DFS}}
=\sum_{\mu,\nu\in\text{DFS}} \rho^{\infty}_{\nu\mu} \bra{\mu}\sigma_x^{(k)}\ket{\nu}
e^{-i(E_\nu-E_\mu)t}.
\end{equation}
As the XX model conserves the total excitation number, each DFS eigenstate $|\mu\rangle$ has a definite excitation number $\mathcal{N}$. Moreover, $\sigma_x^{(k)}=\sigma_+^{(k)}+\sigma_-^{(k)}$ can flip only a single qubit, and consequently this gives rise to the selection rule that unless $\mathcal{N}_\nu=\mathcal{N}_\mu\pm 1$, we have $\bra{\mu}\sigma_x^{(k)}\ket{\nu}=0$. Considering this fact, we are now able to express Eq.~(\ref{expvalue1}) as
\begin{equation} \label{expvalue2}
\langle \sigma_x^{(k)}(t)\rangle_{\text{DFS}}
=\sum_{\substack{\mu,\nu\in\text{DFS}\\ \mathcal{N}_\nu=\mathcal{N}_\mu\pm 1}}
\rho^{\infty}_{\nu\mu}
\bra{\mu}\sigma_x^{(k)}\ket{\nu}
e^{-i(E_\nu-E_\mu)t}.
\end{equation}
A further simplification is obtained by writing Eq.~(\ref{expvalue2}) as the sum of $\mathcal{N}_\nu=\mathcal{N}_\mu + 1$ and $\mathcal{N}_\nu=\mathcal{N}_\mu - 1$ parts. Define 
\begin{equation}
S(t)=\sum_{\substack{\mu,\nu\in\text{DFS}\\ \mathcal{N}_\nu=\mathcal{N}_\mu+1}} 
\rho^{\infty}_{\nu\mu}
\bra{\mu}\sigma_x^{(k)}\ket{\nu}
e^{-i(E_\nu-E_\mu)t}.
\end{equation}
Then, using
$(\rho^{\infty}_{\nu\mu})^*=\rho^{\infty}_{\mu\nu}$ and $\bra{\mu}\sigma_x^{(k)}\ket{\nu}^*=\bra{\nu}\sigma_x^{(k)}\ket{\mu}$, together with $\big(e^{-i(E_\nu-E_\mu)t}\big)^*=e^{-i(E_\mu-E_\nu)t}$, it can be seen that $S(t)^*$ equals the $\mathcal{N}_\nu=\mathcal{N}_\mu-1$ part of Eq.~(\ref{expvalue2}) after swapping the dummy indices $\mu$ and $\nu$. Therefore, using $S(t)+S(t)^*=2\,\text{Re}\,S(t)$ and defining $X_{\mu\nu}^{(k)}=\langle \mu \vert \sigma_x^{(k)} \vert \nu \rangle$, we finally obtain
\vspace{-0.15\baselineskip}
\begin{equation} \label{genexpval}
\langle \sigma_x^{(k)}(t)\rangle_{\text{DFS}}
= 2\mathrm{Re}
\sum_{\substack{\mu,\nu\in\text{DFS}\\ \mathcal{N}_\nu=\mathcal{N}_\mu+1}}
\rho^{\infty}_{\nu\mu}
X_{\mu \nu}^{(k)}
e^{-i(E_\nu-E_\mu)t},
\end{equation}
which is our Eq.~(\ref{eq:genexpval}) in the main text, where we omit writing the $\mathcal{N}_\nu=\mathcal{N}_\mu+1$ condition in the sum for brevity.

\section{Necessary and Sufficient Condition for generic synchronization}

We define generic stable synchronization between the edge qubits by the condition
\begin{equation}
    \langle \sigma_x^{(1)}(t) \rangle_{\text{DFS}}=C\langle \sigma_x^{(N)}(t) \rangle_{\text{DFS}},
\end{equation}
for all $t$, once the dynamics is confined to the DFS, where $C$ is a real constant, and neither asymptotic edge expectation value vanishes at all times. As we demand this condition to hold for arbitrary initial states, the coherences $\rho^\infty_{\nu\mu}$ entering Eq.~(\ref{genexpval}) are generically nonzero, and the relation must therefore be satisfied term by term, i.e.,
\begin{equation}
    X_{\mu\nu}^{(1)}=CX_{\mu\nu}^{(N)},
\end{equation}
for all pairs of DFS eigenstates $\ket{\mu}$ and $\ket{\nu}$ whose excitation numbers differ by one. Since $\sigma_x^{(k)}=\sigma_-^{(k)}+\sigma_+^{(k)}$, in general, both $\mathcal{N}_\nu=\mathcal{N}_\mu\pm1$ transitions contribute to the sum. However, after fixing the ordering $\mathcal{N}_\nu=\mathcal{N}_\mu+1$ in Eq.~(\ref{genexpval}), the relevant matrix elements reduce to those of $\sigma_-^{(k)}$. Let us first calculate these relevant matrix elements 
\begin{align}\label{eq:01me}
\bra{G}\sigma_-^{(1)}\ket{D_{\alpha}}&=\bra{G}\sigma_-^{(1)}\sum_{j=1}^N\phi_{\alpha}(j)\ket{j}=\phi_{\alpha}(1), \nonumber \\
\bra{G}\sigma_-^{(N)}\ket{D_{\alpha}}&=\bra{G}\sigma_-^{(N)}\sum_{j=1}^N\phi_{\alpha}(j)\ket{j}=\phi_{\alpha}(N).
\end{align}
Considering the parity symmetry of the eigenstate amplitudes with respect to the center of the chain, $\phi_n(N + 1 - j) = (-1)^{n+1}\phi_n(j)$, we can conclude that  $\phi_{\alpha}(1)=(-1)^{\alpha+1}\phi_{\alpha}(N)$. This result immediately guarantees that \emph{if} there is exactly one single-excitation DFS state, i.e. $g=\gcd(m, N+1)=2$, qubits at the ends of the chain are synchronized for odd $\alpha$ and anti-synchronized for even $\alpha$, where the unique DFS label is $\alpha=n_1=(N+1)/g=(N+1)/2$, which in turn requires $N$ to be odd. Moreover, due to Eq.~(\ref{genexpval}), the synchronized oscillations occur at the frequency set by the energy difference between the ground state and the unique single-excitation DFS state, namely, $E_{D_\alpha}-E_G=E_{D_\alpha}$.

We now show that stable generic synchronization is achieved \emph{only if} the DFS supports exactly one single-excitation eigenstate, completing the \emph{if and only if} statement. To this end, consider the case of two or more single-excitation DFS states, which allows the construction of at least one two-excitation DFS state via Slater determinants.  In this case, we need the contributions coming from the matrix elements between the single- and double-excitation DFS states. We focus on the matrix elements that contribute to the same oscillatory mode as the $\ket{G}$ and $\ket{D_\alpha}$ transition. Since the energies are additive, $E_{D_{\alpha\beta}}=E_{D_\alpha}+E_{D_\beta}$, the energy difference $E_{D_{\alpha\beta}}-E_{D_\beta}=E_{D_\alpha}$ matches the frequency of the transition between $\ket{G}$ and $\ket{D_\alpha}$. Then, the relevant matrix elements are $\bra{D_\beta}\sigma_-^{(1)}\ket{D_{\alpha\beta}}$ and $\bra{D_\beta}\sigma_-^{(N)}\ket{D_{\alpha\beta}}$. We evaluate these by computing $\sigma_-$ acting on the two-excitation ket explicitly.
\begin{align}
    \sigma_-^{(1)}\ket{D_{\alpha\beta}}&=\sigma_-^{(1)}\sum_{1\leq i<j}^N\left(\phi_{\alpha}(i)\phi_{\beta}(j)-\phi_{\alpha}(j)\phi_{\beta}(i)\right)\ket{i, j}.
\end{align}
The constraint $i<j$ restricts the sum to $i=1$, giving
\begin{align}
    \sigma_-^{(1)}\ket{D_{\alpha\beta}}&=\sum_{j=2}^N\left(\phi_{\alpha}(1)\phi_{\beta}(j)-\phi_{\alpha}(j)\phi_{\beta}(1)\right)\ket{j} \nonumber \\
    &=\phi_{\alpha}(1)\sum_{j=2}^N\phi_{\beta}(j)\ket{j}-\phi_{\beta}(1)\sum_{j=2}^N\phi_{\alpha}(j)\ket{j}.
\end{align}
Using $\ket{D_{\alpha}}=\sum_{j=1}^N\phi_{\alpha}(j)\ket{j}$, we can write $\sum_{j=2}^N\phi_{\alpha}(j)\ket{j}=\ket{D_{\alpha}}-\phi_{\alpha}(1)\ket{1}$, and similarly for the $\beta$ term. Substituting these, we obtain
\begin{align}
\sigma_-^{(1)}\ket{D_{\alpha\beta}}&=\phi_{\alpha}(1)\left(\ket{D_{\beta}}-\phi_{\beta}(1)\ket{1}\right)-\phi_{\beta}(1)\left(\ket{D_{\alpha}}-\phi_{\alpha}(1)\ket{1}\right) \nonumber \\
&=\phi_{\alpha}(1)\ket{D_{\beta}}-\phi_{\beta}(1)\ket{D_{\alpha}}.
\end{align}
Similarly, on the other end of the chain,
\begin{align}
    \sigma_-^{(N)}\ket{D_{\alpha\beta}}&=\sigma_-^{(N)}\sum_{1\leq i<j}^N\left(\phi_{\alpha}(i)\phi_{\beta}(j)-\phi_{\alpha}(j)\phi_{\beta}(i)\right)\ket{i, j} \nonumber \\
    &=\sum_{i=1}^{N-1}\left(\phi_{\alpha}(i)\phi_{\beta}(N)-\phi_{\alpha}(N)\phi_{\beta}(i)\right)\ket{i} \nonumber \\
    &=\phi_{\beta}(N)\sum_{i=1}^{N-1}\phi_{\alpha}(i)\ket{i}-\phi_{\alpha}(N)\sum_{i=1}^{N-1}\phi_{\beta}(i)\ket{i}.
\end{align}
Similarly, using $\sum_{i=1}^{N-1}\phi_{\alpha}(i)\ket{i}=\ket{D_{\alpha}}-\phi_{\alpha}(N)\ket{N}$, and likewise for the $\beta$ term, we obtain
\begin{align}
\sigma_-^{(N)}\ket{D_{\alpha\beta}}&=\phi_{\beta}(N)\left(\ket{D_{\alpha}}-\phi_{\alpha}(N)\ket{N}\right)-\phi_{\alpha}(N)\left(\ket{D_{\beta}}-\phi_{\beta}(N)\ket{N}\right) \nonumber \\
&=-\phi_{\alpha}(N)\ket{D_{\beta}}+\phi_{\beta}(N)\ket{D_{\alpha}}.
\end{align}
Taking the inner product with $\bra{D_\beta}$, we find
\begin{align}\label{eq:12me}
\bra{D_{\beta}}\sigma_-^{(1)}\ket{D_{\alpha\beta}}=\phi_{\alpha}(1)  \qquad  \text{and} \qquad \bra{D_{\beta}}\sigma_-^{(N)}\ket{D_{\alpha\beta}}=-\phi_{\alpha}(N). 
\end{align}
Applying the condition $X_{\mu\nu}^{(1)}=CX_{\mu\nu}^{(N)}$ to Eqs.~(\ref{eq:01me}) and~(\ref{eq:12me}) respectively gives
\begin{align}
    \phi_{\alpha}(1)=C\phi_{\alpha}(N)  \qquad  \text{and}  \qquad  \phi_{\alpha}(1)=-C\phi_{\alpha}(N),
\end{align}
which can only be satisfied if $\phi_{\alpha}(1)=\phi_{\alpha}(N)=0$. This demonstrates that whenever more than one single-excitation DFS state exists, enabling the construction of two-excitation DFS states, generic synchronization is impossible.

Although the argument above rules out \emph{generic} synchronization whenever $g>2$, it does not rule out synchronization altogether. The contradiction between Eqs.~(\ref{eq:01me}) and~(\ref{eq:12me}) only arises when both sets of matrix elements appear in Eq.~(\ref{genexpval}) with non-zero weights $\rho^{\infty}_{\nu\mu}$. For initial states that give $\rho^\infty_{D_\beta D_{\alpha\beta}}=0$ for all $\beta$, the problematic terms simply drop out, leaving only the $\bra{G}\sigma_-\ket{D_\alpha}$ contributions at frequency $E_{D_\alpha}$, and the synchronization behavior is governed by the single condition $X_{GD_\alpha}^{(1)}=CX_{GD_\alpha}^{(N)}$. This is indeed exactly what happens in the case study of Fig.~\hyperref[fig:Fig1]{\ref*{fig:Fig1}(a)}. The initial state lives entirely in the zero- and single-excitation sectors, so coherences involving multi-excitation DFS states never build up, and the edge qubits end up synchronizing at multiple frequencies set by the single-excitation DFS energies. In fact, any initial state whose support restricts $\rho^\infty$ to a compatible subset of DFS coherences can still give rise to synchronization, but its frequency components and $C$ will depend on that particular choice of state.

\section{Entanglement between the edge qubits}

In any noise setting that satisfies the condition $g=\gcd(m_1,\dots,m_q,N+1)=2$, the only surviving single-excitation DFS state is $\ket{D_{(N+1)/2}}\equiv\ket{D}$ with energy $E_{D}=\omega+2J\cos(\pi/2)=\omega$. The DFS then consists of just two states: the ground state $\ket{G}$ and the single-excitation state $\ket{D}$, where $E_D-E_G=\omega$. Using Eq.~(\ref{rhodfst}), the time-dependent density operator within the DFS takes the form
\begin{equation}
\rho_{\text{DFS}}(t) = \rho_{GG}^{\infty} |G\rangle\langle G| + \rho_{DD}^{\infty} |D\rangle\langle D| + \rho_{DG}^{\infty} e^{-i\omega t} |D\rangle\langle G| + \rho_{GD}^{\infty} e^{+i\omega t} |G\rangle\langle D|,
\end{equation}
where $\rho_{\bullet\bullet}^{\infty}=\bra{\bullet}\rho^{\infty}\ket{\bullet}$ are the matrix elements of the asymptotic density operator $\rho^{\infty}$ introduced in Eq.~(\ref{rhodfst}), with $\rho_{GD}^{\infty} = (\rho_{DG}^{\infty})^*$ and $\rho_{GG}^{\infty}+\rho_{DD}^{\infty}=1$. To obtain the reduced density matrix of the edge qubits, we trace over all bulk sites $j=2,\ldots,N-1$. The diagonal term $\ket{G}\bra{G}$ has all qubits in $\ket{0}$ and contributes $\ket{00}\bra{00}$. For $\ket{D}\bra{D}=\sum_{i,j}\phi(i)\phi(j)\ket{i}\bra{j}$, the partial trace retains only terms where the bulk configurations match on both sides. When the excitation is on an edge site, that is $i,j\in\{1,N\}$, the bulk is in $\ket{0\cdots0}$ on both sides and the trace gives
\begin{equation}
\phi^2(1)\ket{10}\bra{10}+\phi^2(N)\ket{01}\bra{01}+\phi(1)\phi(N)\big(\ket{10}\bra{01}+\ket{01}\bra{10}\big).
\end{equation}
When the excitation is on a bulk site, that is $i=j\in\{2,\ldots,N-1\}$, both edge qubits are in $\ket{0}$, contributing
\begin{equation}
\sum_{j=2}^{N-1}\phi^2(j)\,\ket{00}\bra{00} = \big[1-\phi^2(1)-\phi^2(N)\big]\ket{00}\bra{00},
\end{equation}
where we used the normalization $\sum_{j=1}^N\phi^2(j)=1$. Cross terms with one edge and one bulk index vanish since the bulk configurations differ. For the off-diagonal term $\ket{D}\bra{G}$, the bra $\bra{G}$ fixes all bulk qubits to $\ket{0}$, so only edge-site components of $\ket{D}$ survive as $\phi(1)\ket{10}\bra{00}+\phi(N)\ket{01}\bra{00}$. The $\ket{G}\bra{D}$ term is obtained by Hermitian conjugation. Combining all the terms above, the reduced density matrix of the edge qubits is given by
\begin{align}\label{eq:rho1N}
    \rho_{\text{DFS}}^{(1,N)}(t) &= \left[\rho_{GG}^{\infty}+\rho_{DD}^{\infty}\left(1-\phi^2(1)-\phi^2(N)\right)\right] |00\rangle\langle 00| \nonumber \\ 
    &+\rho_{DD}^{\infty}\left[\phi^2(1)|10\rangle\langle 10|+\phi^2(N)|01\rangle\langle 01|+\phi(1)\phi(N)\left(|10\rangle\langle 01|+|01\rangle\langle 10|\right)\right] \nonumber \\
    &+\rho_{DG}^{\infty}e^{-i\omega t}\left[\phi(1)|10\rangle\langle 00|+\phi(N)|01\rangle\langle 00|\right] \nonumber \\
    &+\rho_{GD}^{\infty} e^{+i\omega t}\left[\phi(1)|00\rangle\langle 10|+\phi(N)|00\rangle\langle 01|\right],
\end{align}
where $\phi(1)=\sqrt{2/(N+1)}$, and $\phi(N)=\sqrt{2/(N+1)}\sin(\pi N/2)=s\sqrt{2/(N+1)}$ with $s=(-1)^{\frac{N-1}{2}}$. Expressed in the computational basis $\{\ket{00},\ket{01},\ket{10},\ket{11}\}$, the reduced density operator takes the form
\begin{equation}\label{eq:rho1N_matrix}
\rho_{\text{DFS}}^{(1,N)}(t) = 
\begin{pmatrix} 
\rho_{GG}^{\infty}+\rho_{DD}^{\infty}\left(\frac{N-3}{N+1}\right) & s\sqrt{\frac{2}{N+1}}\rho_{GD}^{\infty}e^{i\omega t} & \sqrt{\frac{2}{N+1}}\rho_{GD}^{\infty}e^{i\omega t} & 0 \\[7pt] 
s\sqrt{\frac{2}{N+1}}\rho_{DG}^{\infty}e^{-i\omega t} & \frac{2}{N+1}\rho_{DD}^{\infty} & s\frac{2}{N+1}\rho_{DD}^{\infty} & 0 \\[7pt] 
\sqrt{\frac{2}{N+1}}\rho_{DG}^{\infty}e^{-i\omega t} & s\frac{2}{N+1}\rho_{DD}^{\infty} & \frac{2}{N+1}\rho_{DD}^{\infty} & 0 \\[7pt] 
0 & 0 & 0 & 0 
\end{pmatrix}.
\end{equation}
Note that the $\ket{11}$ row and column vanish since neither $\ket{G}$ nor $\ket{D}$ contains a double excitation on the edge sites.

The concurrence of the edge qubits can be calculated from~\mbox{[S1]}
\begin{equation}
\mathcal{C}=\max\{0,\lambda_1-\lambda_2-\lambda_3-\lambda_4\},
\end{equation}
where $\lambda_i$ are the square roots of the eigenvalues of $\rho\tilde{\rho}$ in decreasing order, with $\tilde{\rho}=(\sigma_y\otimes\sigma_y)\rho^*(\sigma_y\otimes\sigma_y)$. A direct calculation of $\rho\tilde{\rho}$ shows that it has only one nonzero eigenvalue so that
\begin{equation}\label{eq:concurrence}
\mathcal{C}^\infty_{(1,N)}= 2\abs{s\frac{2}{N+1}\rho_{DD}^{\infty}} = \frac{4\abs{\rho_{DD}^{\infty}}}{N+1},
\end{equation}
which is clearly time-independent. This establishes that when generic synchronization is present, i.e., $g=2$, constant asymptotic entanglement between the edge qubits is guaranteed.

Lastly, we determine $\rho_{DD}^{\infty}$ for the initial state in our case study, i.e., for the states of the form $\ket{\psi}=\ket{+}_1\otimes\bigotimes_{k=2}^{N}\ket{0}_k$. This state resides within the zero- and single-excitation sectors, ensuring no population can flow in from higher-excitation manifolds during the transient evolution. The overlap with the DFS eigenstate is
\begin{equation}
\rho_{DD}^{\infty}=\abs{\braket{D}{\psi}}^2=\phi^2(1)\,\abs{\braket{1}{+}}^2=\left(\frac{2}{N+1}\right) \frac{1}{2}=\frac{1}{N+1}.
\end{equation}
Substituting into Eq.~(\ref{eq:concurrence}), the concurrence reads
\begin{equation}
\mathcal{C}^{\infty}_{(1,N)}=\frac{4}{(N+1)^2}.
\end{equation}

\vspace{1em}
\noindent [S1] W. K. Wootters, Phys. Rev. Lett. \textbf{80}, 2245 (1998).

\end{document}